\documentstyle[twoside]{article}
\input epsf.tex

\catcode`\@=11
\long\def\@makefntext#1{
\protect\noindent \hbox to 3.2pt {\hskip-.9pt  
$^{{\eightrm\@thefnmark}}$\hfil}#1\hfill}		

\def\@makefnmark{\hbox to 0pt{$^{\@thefnmark}$\hss}}	
	
\def\ps@myheadings{\let\@mkboth\@gobbletwo
\def\@oddhead{\hbox{}
\rightmark\hfil\eightrm\thepage}   
\def\@oddfoot{}\def\@evenhead{\eightrm\thepage\hfil
\leftmark\hbox{}}\def\@evenfoot{}
\def\sectionmark##1{}\def\subsectionmark##1{}}

\oddsidemargin=\evensidemargin
\newcounter{sectionc}\newcounter{subsectionc}\newcounter{subsubsectionc}
\renewcommand{\section}[1] {\vspace{12pt}\addtocounter{sectionc}{1} 
\setcounter{subsectionc}{0}\setcounter{subsubsectionc}{0}\noindent 
	{\tenbf\thesectionc. #1}\par\vspace{5pt}}
\renewcommand{\subsection}[1] {\vspace{12pt}\addtocounter{subsectionc}{1} 
	\setcounter{subsubsectionc}{0}\noindent 
	{\bf\thesectionc.\thesubsectionc. {\kern1pt \bfit #1}}\par\vspace{5pt}}
\renewcommand{\subsubsection}[1] {\vspace{12pt}\addtocounter{subsubsectionc}{1}
	\noindent{\tenrm\thesectionc.\thesubsectionc.\thesubsubsectionc.
	{\kern1pt \tenit #1}}\par\vspace{5pt}}

\newcounter{appendixc}
\newcounter{subappendixc}[appendixc]
\newcounter{subsubappendixc}[subappendixc]
\renewcommand{\thesubappendixc}{\Alph{appendixc}.\arabic{subappendixc}}
\renewcommand{\thesubsubappendixc}
	{\Alph{appendixc}.\arabic{subappendixc}.\arabic{subsubappendixc}}

\renewcommand{\appendix}[1] {\vspace{12pt}
        \refstepcounter{appendixc}
        \setcounter{figure}{0}
        \setcounter{table}{0}
        \setcounter{lemma}{0}
        \setcounter{theorem}{0}
        \setcounter{corollary}{0}
        \setcounter{definition}{0}
        \setcounter{equation}{0}
        \renewcommand{\thefigure}{\Alph{appendixc}.\arabic{figure}}
        \renewcommand{\thetable}{\Alph{appendixc}.\arabic{table}}
        \renewcommand{\theappendixc}{\Alph{appendixc}}
        \renewcommand{\thelemma}{\Alph{appendixc}.\arabic{lemma}}
        \renewcommand{\thetheorem}{\Alph{appendixc}.\arabic{theorem}}
        \renewcommand{\thedefinition}{\Alph{appendixc}.\arabic{definition}}
        \renewcommand{\thecorollary}{\Alph{appendixc}.\arabic{corollary}}
        \renewcommand{\theequation}{\Alph{appendixc}.\arabic{equation}}
        \noindent{\tenbf Appendix \theappendixc #1}\par\vspace{5pt}}
\newcommand{\subappendix}[1] {\vspace{12pt}
        \refstepcounter{subappendixc}
        \noindent{\bf Appendix \thesubappendixc. {\kern1pt \bfit #1}}
	\par\vspace{5pt}}
\newcommand{\subsubappendix}[1] {\vspace{12pt}
        \refstepcounter{subsubappendixc}
        \noindent{\rm Appendix \thesubsubappendixc. {\kern1pt \tenit #1}}
	\par\vspace{5pt}}

\newcommand{\textlineskip}{\baselineskip=13pt}
\newcommand{\smalllineskip}{\baselineskip=10pt}

\def\eightcirc{
\begin{picture}(0,0)
\put(4.4,1.8){\circle{6.5}}
\end{picture}}
\def\eightcopyright{\eightcirc\kern2.7pt\hbox{\eightrm c}}

\def\abstracts#1#2#3{{
	\centering{\begin{minipage}{4.5in}\baselineskip=10pt\footnotesize
	\parindent=0pt #1\par 
	\parindent=15pt #2\par
	\parindent=15pt #3
	\end{minipage}}\par}} 

\renewenvironment{thebibliography}[1]
	{\frenchspacing
	 \ninerm\baselineskip=11pt
	 \begin{list}{\arabic{enumi}.}
	{\usecounter{enumi}\setlength{\parsep}{0pt}
	 \setlength{\leftmargin 12.7pt}{\rightmargin 0pt} 
	 \setlength{\itemsep}{0pt} \settowidth
	{\labelwidth}{#1.}\sloppy}}{\end{list}}

\newcounter{itemlistc}
\newcounter{romanlistc}
\newcounter{alphlistc}
\newcounter{arabiclistc}

\newcommand{\fcaption}[1]{
        \refstepcounter{figure}
        \setbox\@tempboxa = \hbox{\footnotesize Fig.~\thefigure. #1}
        \ifdim \wd\@tempboxa > 5in
           {\begin{center}
        \parbox{5in}{\footnotesize\smalllineskip Fig.~\thefigure. #1}
            \end{center}}
        \else
             {\begin{center}
             {\footnotesize Fig.~\thefigure. #1}
              \end{center}}
        \fi}

\newcommand{\tcaption}[1]{
        \refstepcounter{table}
        \setbox\@tempboxa = \hbox{\footnotesize Table~\thetable. #1}
        \ifdim \wd\@tempboxa > 5in
           {\begin{center}
        \parbox{5in}{\footnotesize\smalllineskip Table~\thetable. #1}
            \end{center}}
        \else
             {\begin{center}
             {\footnotesize Table~\thetable. #1}
              \end{center}}
        \fi}

\def\@citex[#1]#2{\if@filesw\immediate\write\@auxout
	{\string\citation{#2}}\fi
\def\@citea{}\@cite{\@for\@citeb:=#2\do
	{\@citea\def\@citea{,}\@ifundefined
	{b@\@citeb}{{\bf ?}\@warning
	{Citation `\@citeb' on page \thepage \space undefined}}
	{\csname b@\@citeb\endcsname}}}{#1}}

\newif\if@cghi
\def\cite{\@cghitrue\@ifnextchar [{\@tempswatrue
	\@citex}{\@tempswafalse\@citex[]}}
\def\citelow{\@cghifalse\@ifnextchar [{\@tempswatrue
	\@citex}{\@tempswafalse\@citex[]}}
\def\@cite#1#2{{$\null^{#1}$\if@tempswa\typeout
	{IJCGA warning: optional citation argument 
	ignored: `#2'} \fi}}

\def\pmb#1{\setbox0=\hbox{#1}
	\kern-.025em\copy0\kern-\wd0
	\kern.05em\copy0\kern-\wd0
	\kern-.025em\raise.0433em\box0}

\def\fnt#1#2{\footnotetext{\kern-.3em
	{$^{\mbox{\scriptsize #1}}$}{#2}}}

\def\fpage#1{\begingroup
\voffset=.3in
\thispagestyle{empty}\begin{table}[b]\centerline{\footnotesize #1}
	\end{table}\endgroup}

\font\tenrm=cmr10
\font\tenit=cmti10 
\font\tenbf=cmbx10
\font\bfit=cmbxti10 at 10pt
\font\ninerm=cmr9

\font\eightrm=cmr8

\def\qed{\hbox{${\vcenter{\vbox{			
   \hrule height 0.4pt\hbox{\vrule width 0.4pt height 6pt
   \kern5pt\vrule width 0.4pt}\hrule height 0.4pt}}}$}}

\newcommand{\be}{\begin{equation}}   \newcommand{\ee}{\end{equation}}
\newcommand{\bear}{\begin{eqnarray}}
\newcommand{\eear}{\end{eqnarray}}
\newcommand{\ba}{\begin{array}}      \newcommand{\ea}{\end{array}}

\newcommand{\CQ}{{\cal Q}}
\newcommand{\CU}{{\cal U}}
\newcommand{\CD}{{\cal D}}
\newcommand{\CL}{{\cal L}}
\newcommand{\CE}{{\cal E}}

\newcommand{\ov}{\overline}

\begin{document}


\normalsize\textlineskip
\thispagestyle{empty}
\setcounter{page}{1}


\vspace*{-1.5cm}
\begin{flushright}
EFI-2000-43
\end{flushright}
\vspace*{-1.2cm}

\vspace*{0.88truein}

\fpage{1}
\centerline{\bf MINIMAL ELECTROWEAK SYMMETRY BREAKING MODEL}
\vspace*{0.035truein}
\centerline{\bf IN EXTRA DIMENSIONS
}
\vspace*{0.37truein}
\centerline{\footnotesize HSIN-CHIA CHENG
}
\vspace*{0.015truein}
\centerline{\footnotesize\it Enrico Fermi Institute, The University of 
Chicago}
\baselineskip=10pt
\centerline{\footnotesize\it Chicago, IL 60637, USA
}

\vspace*{0.21truein}
\abstracts{We show that if the Standard Model gauge fields and fermions
propagate in extra dimenions, a composite Higgs field with the correct
quantum number can arise naturally as a bound state due to the strong
gauge interactions in higher dimensions. The top quark mass and the Higgs
mass can be predicted from the infrared fixed points of the renormalization
group equations. The top quark mass is in good agreement with the 
experimental value, and the Higgs boson mass is predicted to be
$\sim 200$GeV. There may be some other light bound states which could be
observed at upcoming collider experiments.
}{}{}

\vspace*{12pt}			

\vspace*{1pt}\textlineskip	

The origin of the electroweak symmetry breaking (EWSB) is currently one of
the most important questions in particle physics. In the Standard Model 
(SM), it is achieved by a nonzero vacuum expectation value
of a fundamental 
scalar Higgs field. However, this picture is not totally 
satisfactory: there is no understanding of the
gauge quantum numbers of the Higgs field, and why its squared mass
is negative. In addition, the squared-mass of a fundamental scalar
field receives quadratically divergent radiative
corrections, hence suffers from the ``hierarchy problem'' if the
cutoff scale is much higher than the weak scale. This problem can be
avoided if the Higgs is a composite object rather than a fundamental
field, and cease to be a dynamical degree of freedom not much above
the weak scale.
In fact, even without
the Higgs field, the electroweak symmetry would still be broken by
the quark-anti-quark condensate when QCD becomes strong at low energies.
The breaking scale is three orders of magnitude too small.
Nevertheless, it shows that a composite Higgs condensate
can arise if there is strong interaction at the TeV scale.

Gauge interactions in more
than four dimensions are non-renormalizable and rapidly become strong
at high energies. Therefore, 
the SM gauge interactions can become strong at the TeV scale
and be responsible for forming the composite Higgs 
if they propagate in extra dimensions of the TeV$^{-1}$ size. Furthermore,
the constituents of the composite Higgs field can naturally be the
SM top quark and its Kaluza-Klein (KK) excitations.\cite{CDH,ACDH}
The model is minimal in the sense that no new fields or interactions 
beyond the observed in the Standard Model are needed for the EWSB.
It can also give powerful predictions which are in
good agreement with the experimental results. We will discuss an example
in the following.

Consider a one generation model model in which the SM gauge fields 
and the third generation
fermions live in six dimensions, with two of the six dimensions compactified
at a scale $M_c \sim {\rm TeV}^{-1}$. (Inclusion of the first two generations
will be discussed later.) The theory is non-renormalizable hence needs
a physical cutoff $M_s$. A possible candidate is the scale of quantum gravity,
which is determined by the sizes of the extra dimensions accessible to the 
gravitons.\cite{ADD,RS} 
In six dimensions, there exist four-component chiral fermions.
We assign $SU(2)_W$ doublets with positive chirality, $\CQ_+$,
$\CL_+$, and $SU(2)_W$ singlets with negative chirality, $\CU_-$,
$\CD_-$, $\CE_-$. Each fermion contains both left- and right-handed
two-component spinors when reduced to four dimensions. We impose an
orbifold projection such that the  right-handed components of $\CQ_+$,
$\CL_+$, and left-handed components of $\CU_-$, $\CD_-$, $\CE_-$, are
odd under the orbifold ${\bf Z}_2$ symmetry and therefore the
corresponding zero modes are projected out.  As a result, the
zero-mode fermions are two-component four-dimensional  quarks and
leptons: ${\cal Q}_+^{(0)} \equiv (t,b)_L$, \  ${\cal U}_-^{(0)}
\equiv t_R$, \  ${\cal D}_-^{(0)} \equiv b_R$, \  ${\cal L}_+^{(0)}
\equiv (\nu_\tau, \tau)_L$,  ${\cal E}_-^{(0)} \equiv \tau_R$.

At the cutoff scale, the SM gauge interactions are strong and will
produce bound states. The squared-mass of a scalar bound state
has quadratic dependence on the cutoff, and can become much smaller than
the cutoff scale or even negative if the coupling is sufficiently strong. 
Using the one-gauge-boson-exchange approximation, one finds in general 
among possible scalar bound states, $H_{\CU}=\ov{\CQ}_+ \CU_-$, which
has the correct quantum number to be the Higgs field, is the
most attractive channel. Therefore it is most likely to aquire a negative
squared-mass to break the electroweak symmetry. The composite Higgs
is expected to have large coupling to its constituents, so it not only
predicts the correct Higgs quantum number, but also a heavy up-type
quark (top quark). The $H_{\CD}= \ov{\CQ}_+ \CU_-$ channel is also quite
strongly bound while the other channels are not sufficiently strong
to produce light bound states. The low-energy theory below $M_c$
is expected to be a two-Higgs-doublet model.

Compared with the usual four-dimensional dynamical EWSB
models, the higher-dimensional model has the advantage that
the binding force can be the SM gauge interactions themselves,
without the need of introducing new strong interactions. In addition,
it also gives a prediction
of the top quark mass naturally in the right range. In the minimal
four-dimensional top quark condensate model, the top quark is too
heavy, $\sim 600$ GeV, 
if the compositeness scale is in the TeV range.\cite{BHL}
The correct top mass can be obtained if some additional vector-like
quarks which also participate in the electroweak symmetry 
breaking.\cite{topseesaw} They are naturally provided by 
the KK excitations of the top quark in a theory with 
extra dimensions. Another way of understanding of the top Yukawa
coupling being $\sim 1$ instead of the strong coupling value $\sim 4\pi$
is that (the zero mode of) the top quark coupling receives a volume
dilution factor because it propagates in extra dimensions.
In fact, the top quark mass can be predicted quite insensitively to
the cutoff because of the infrared fixed point behavior of the 
renormalization group (RG) evalution. The infrared fixed point is
rapidly approached due to the power-law running in extra-dimensional
theories,\cite{DDG} even though the cutoff scale is not much higher
the the weak scale. Similarly, the Higgs self-coupling also receives 
the extra-dimensional volume suppression. As a result, the physical
Higgs boson is relatively light, $\sim 200$ GeV, in contrast with the 
usual dynamical EWSB models. It is also governed
by the infrared fixed point of the RG equations. The numerical predictions
of the top quark mass and the Higgs boson mass are shown in Fig.~1.
\begin{figure}[htbp]
\centerline{\epsfysize=5.9cm\epsfbox{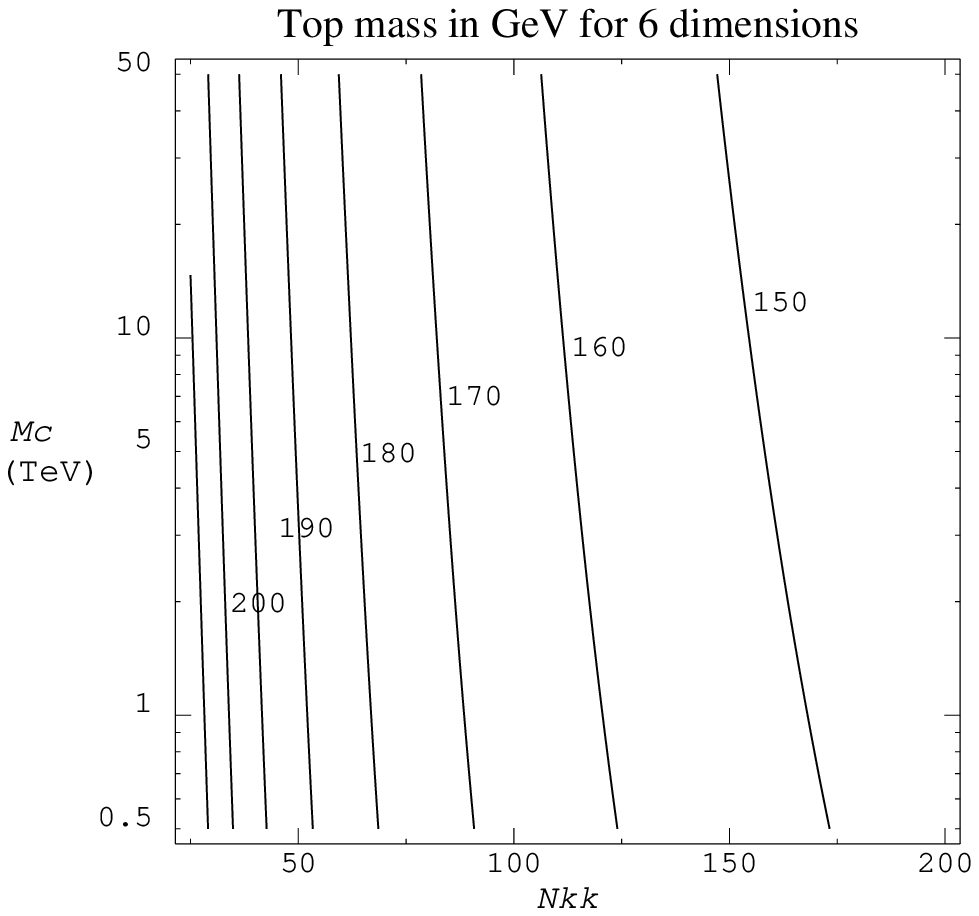}
\epsfysize=5.9cm\epsfbox{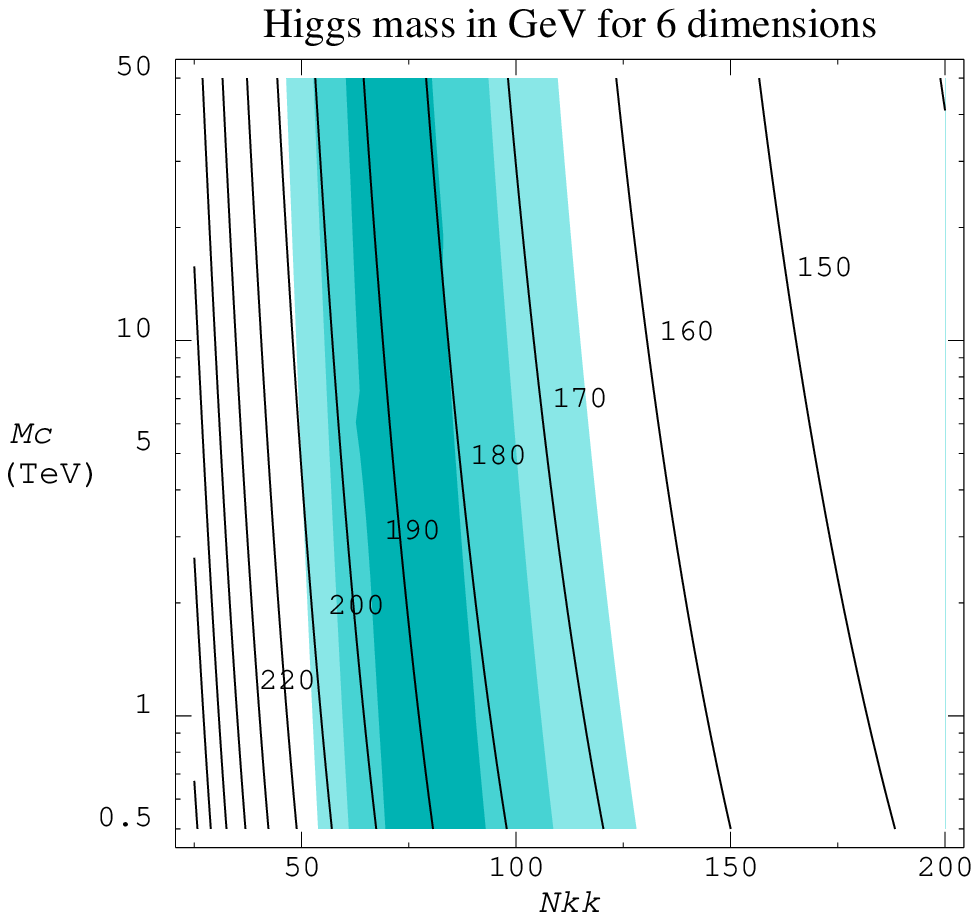}}
\fcaption{The top quark mass (left) and the Higgs boson mass (right)
as functions of the number of KK excitations, $N_{\rm KK}$, and the 
compactification scale $M_c$ in the six-dimensional theory. The shaded
area in the Higgs boson mass prediction corresponds to the top quark
mass lying within $3\sigma$ of the experimental value, $174.3\pm 5.1$~GeV.}
\end{figure}

In more general models, there may be other light bound states 
which could be observed in the upcoming collider experiments
in addition to the Higgs bosons. 
The possible light bound states depend on the model. For example,
in the eight-dimensional model, there is a strongly bound state 
$\ov{\CQ} \CQ^c$ transforming like the right-handed bottom quark
under the SM gauge group.\cite{ACDH}
Inclusion of the first two generations may also introduce more
bound states. If the first two generations are localized in four
dimensions, they can form four-dimensional bound states, though
their binding strength may be smaller because they have no contribution
from the extra components of the gauge fields. If all three generations
of fermions live in the bulk, then we need to introduce explicit
flavor-breaking interactions from the cutoff scale to distinguish
them. The flavor-breaking interactions should enhance
the third generation channels relative to the first two generation
channels so that only the composite Higgs field from the third generation
top quark have a negative squared-mass and is responsible for the
EWSB. These flavor-breaking interactions
can also give masses to the other light fermions.


\end{document}